# A Novel SOC Estimation for Hybrid Energy Pack using Deep Learning

Chigozie Uzochukwu Udeogu
*Department of Aeronautics, Mechanical and Electronic Convergence Engineering*
*Kumoh National Institute of Technology, 39177 Gumi-si, South Korea*
Email: chigozie.udeogu@kumoh.ac.kr

*Abstract*— Estimating the state of charge (SOC) of compound energy storage devices in the hybrid energy storage system (HESS) of electric vehicles (EVs) is vital in improving the performance of the EV. The complex and variable charging and discharging current of EVs makes an accurate SOC estimation a challenge. This paper proposes a novel deep learning-based SOC estimation method for lithium-ion battery-supercapacitor HESS EV based on the nonlinear autoregressive with exogenous inputs neural network (NARXNN). The NARXNN is utilized to capture and overcome the complex nonlinear behaviors of lithium-ion batteries and supercapacitors in EVs. The results show that the proposed method improved the SOC estimation accuracy by 91.5% on average with error values below 0.1% and reduced consumption time by 11.4%. Hence validating both the effectiveness and robustness of the proposed method.

## I. Introduction

LITHIUM-ION batteries have been typically used as a single energy storage system (ESS) in electric vehicles (EVs) and hybrid electric vehicles (HEVs). This is because lithium-ion batteries have high specific energy and are environmentally friendly. However, they have limited specific power and this affects the performance of EVs. The urgent need in energy storage to meet the increasingly serious demands for sustainable energy has given rise to hybrid energy storage systems (HESS) and the supercapacitor (SC) is one of the significant contributing technology [1]. Although SCs have low specific energy, they are also known to have high specific power and exceptionally high cycle life. To improve the operation of EVs, SCs are integrated with lithium-ion batteries to make HESS take advantage of the complementary properties of the two devices [2]. To effectively manage the load demand split between the battery packs and SC for the optimal performance of the HEVs, the energy management system (EMS) was introduced [3-4]. Furthermore, the basis of a rational energy management of HESS is the state of charge (SOC) information of its constituent energy storage systems. Thus, accurate SOC estimation plays a key role in improving the dynamic performance and range of HEVs [5-7].

Several approaches have been proposed to improve SOC estimation accuracy in HEVs. Model-based methods such as the electrochemical model and the equivalent circuit model (ECM) have been used by some researchers [8]. [8] proposed a co-estimator to simultaneously estimate the model parameters and SOC of lithium-ion battery and SC. Although the model had a high estimation accuracy, it suffered from the lack of theoretical validation of the estimation error convergence. Kalman filter (KF)-based methods have also been used in identifying and estimating SOC [9-11]. [11] used KF to accurately describe the dynamic voltage behavior of SC and battery under different load conditions and reliably identify the model parameters. However, efforts aimed at improving SOC estimation accuracy in the model increased the complexity and computation time, hence weakening the improvement effects. Another estimation method used in SOC estimation is the sliding mode observer. [12] proposed a multiobserver-based estimation scheme using sliding mode observers for combined state and parameter estimation of battery-double-layer capacitor (DLC) HESS. A limitation of the proposed scheme is that the performance degenerates under high levels of voltage measurement noise and thermal model uncertainty. Generally, a major drawback of the above approaches is the high computational burden coupled with the accuracy degradation under certain factors.

To cope with the challenges of the methods, artificial neural network (ANN) model is suited for modeling complex non-linear systems and it does not depend on the battery/SC models and mathematical relationship [13-14]. However, existing ANN methods are characterized by slow convergence, data overfitting, and can be easily trapped in a local minimum. The nonlinear autoregressive with exogenous inputs (NARX) model is a dynamic neural network with a feedback unit and a recursive neural network. This structure makes it ideal for predicting complex and nonlinear systems like that of the battery and SC. Also, considering the evolution of SOC of battery/SC discharging at constant current, the model becomes a viable solution. This paper, therefore, propose a NARX-based SOC estimation scheme for lithium-ion battery and SC in HESS. The proposed technique aims to provide a good balance between complexity and accuracy as NARX has a relatively strong learning ability and a better time-series prediction performance.

The contributions of this paper are as follows:
1) A deep learning model-based battery and SC SOC estimation scheme for HESS applications that seeks to improve estimation accuracy while reducing complexity.
2) Experiments are carried out to explore the robustness and practicability of the proposed scheme in an HEV.

To the best knowledge of the author, this is a novel attempt to use artificial intelligence approach especially deep learning models in SOC estimation of lithium-ion battery-SC HESS research of HEV as there seem to be lacking in previous studies an application of deep learning methods to the SOC estimation of lithium-ion battery-SC HESS HEV specifically.

## II. Proposed System

The proposed system is illustrated in Fig. 1. It includes the gathering of measured data to collect voltage and current raw data, data preprocessing and normalization to format collected data, model training, and SOC estimation achieved with NARX; denoted with steps 1,2,3, and 4 for both SC and battery.

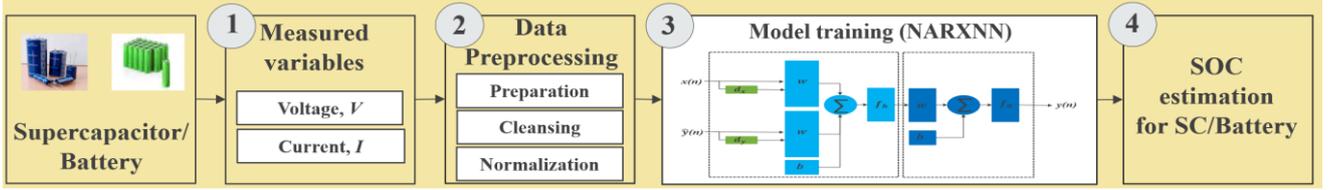

Fig. 1. Flowchart of the proposed SOC estimation for HESS.

*A. System overview*

In the proposed SOC estimation model as shown in Fig. 1, the measured voltage and current data of the SC/battery which are associated with the actual SOC are collected separately in step 1. These collected data are used separately. For each energy storage device – SC and battery, its data are used as input variables of the proposed model. The collected raw data are transformed to a useful and efficient format through data preprocessing in step 2. Data preprocessing involves three parts: preparation, cleansing, and normalization. During preparation, collected capacity data is used to compute the actual SOC. The obtained actual SOC is used in training and testing the SOC estimation. Data cleansing is done to remove outliers. Normalization is applied because an appropriate data normalization can cause the training process of NARXNN to become more efficient and robust. It can also improve the network convergence rate hence reducing complexity. After data preprocessing, the model is trained to estimate the SOC. In step 3, current and voltage data are chosen as input training data while actual SOC data is chosen as target data. In training, the model is trained with charging and discharging data. For the case of battery in step 3, the SOC estimation model is trained with charging and discharging data separately. This is because the charge and discharge of batteries is much more nonlinear when compared to that of SC. At the end of the process, the result of charging and discharging is combined to have a full-cycle result. This technique enables the model to reduce the error in estimation due to the dynamic profile of the battery charging and discharging processes.

*B. SOC estimation using NARX neural network*

SOC is defined as the ratio of the current capacity and nominal capacity of the battery/SC. The nominal capacity represents the maximum charge of the battery/SC. To obtain the SOC mathematically, the definition of SOC can be represented as:

$$SOC(t) = \frac{C_n - \int_0^t I(t)dt}{C_n} \quad (1)$$

where $I(t)$ is the battery/SC current, and $C_n$ is the nominal capacity/capacitance of the battery/SC. The integration of the currents commences at $SOC(t) = 100\%$ when $t = 0$ during discharge.

NARX is a subclass of recurrent neural networks (RNNs).

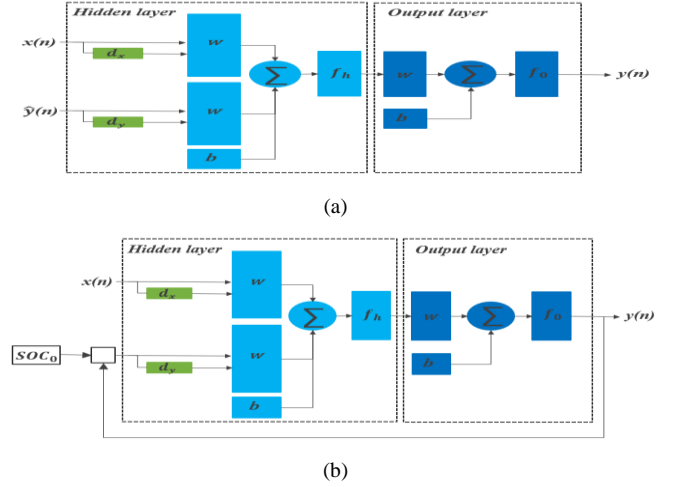

Fig. 2. NARX network architecture. (a) Open-loop configuration for training. (b) Closed-loop configuration for estimation.

It is represented by a discrete nonlinear model and as such is suitable for predicting nonlinear time series problems. NARXNN utilizes limited feedback to form output layers instead of hidden layers. The mathematical expression of the network is defined as:

$$y(n) = f[y(n-1), y(n-2), \ldots, y(n-d_y); x(n-1), x(n-2), \ldots, x(n-d_x)] \quad (2)$$

where $x(n)$ and $y(n)$ denote the network inputs and outputs at discrete time step $n$ respectively, $d_x$ and $d_y$ denote the input and output delay coefficient used in the model respectively, and $f(\cdot)$ denotes the nonlinear function of the network. During NARX regression process, the next value of the successive value of the dependent output signal $y(n)$ is regressed on the preceding $d_y$ values of the output signal and preceding $d_x$ values of the independent (exogenous) input signal.

There are two configuration modes of the NARXNN- the open-loop configuration and the closed-loop configuration. Figure 2 shows the two configuration modes of the NARXNN structure. In this proposed model, NARX open-loop configuration mode is adopted during the training phase, while the closed-loop mode is utilized during the estimation stage. The output regressor of the open-loop mode of the network can be expressed mathematically as:

$$y(n) = f[\hat{y}(n-1), \hat{y}(n-2), \ldots, \hat{y}(n-d_y); x(n-1), x(n-2), \ldots, x(n-d_x)] \quad (3)$$

Here, utilizing the actual output as target, a supervised training technique is carried out. This is desirable because the input of the network is more accurate and the resultant network has a feedforward structure. The output estimation of the closed-loop mode is mathematically expressed as (2). When switching to the closed-loop mode, the estimation has an unknown value and this cannot be fed back to the network as it will cause a divergence in the estimation over time. To mitigate this, a constant value $SOC_0$ is used to replace the feedback signal at the first second of the computation via a time delay. $SOC_0$ is the last SOC value recorded in the previous step. After the initial second of computation, when it is observed that the output estimation has gained stability, the input switches back to the feedback signal. The time instant, $n_0$ when the feedback signal switches from $SOC_0$ to the estimated output, the equation describing the model can be expressed as:

$$y(n) = f[SOC_0; x(n-1, x((n-2), \ldots, x(n-d_x))], \quad n < n_0 \quad (4)$$

and

$$y(n) = f(y(n-1), y(n-2), \ldots, y(n-d_y); x(n-1), x(n-2), \ldots, x(n-d_x)], \quad n \geq n_0. \quad (5)$$

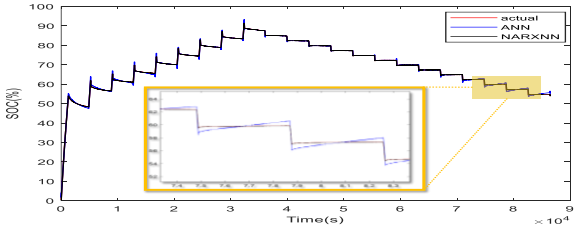

Fig. 3. SOC estimation results of the SC under CC+CV and an example interval (insert).

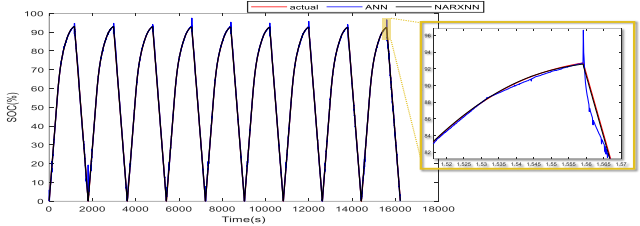

Fig. 4. SOC estimation results of the battery under CC+CV and an example interval (insert).

This paper considers current and voltage as inputs and SOC as the output of the NARXNN since SOC is closely related to the current and terminal voltage of the battery/SC. Hence, a three-layer NARXNN is considered where the battery/SC current and terminal voltage are fed as input vectors and SOC obtained at the same sampling time from the charging and discharging experiment of the battery/SC is the network output. With the aid of self-learning, the parameters identification for the NARXNN is determined by the training stage of the network. Here, the NARXNN open-loop mode was adopted in training the network. The training set data decides the NARXNN with optimal parameters. To validate the performance ability of the proposed model in estimating SOC, the test set data is used during the test stage of NARXNN. During the testing stage, the target values are unknown, thus the open-loop training mode of NARXNN is transformed into a closed-loop mode where the predicted SOC values are backward to the input with time delay. The simulation was carried out in MATLAB where 70 percent of the data was used to train the network, 15 percent was used for validation, and 15 percent was used for testing. The Levenberg-Marquart learning algorithm which is one of the fastest algorithms for adjusting weights for a neural network was used. The open-loop training mode of the NARXNN was obtained with 1:2 defined as the input and output time-delay order and 16 neurons set in the hidden layer.

TABLE. 1. PERFORMANCE OF SOC ESTIMATION UNDER CC+CV PROFILE AT ROOM TEMPERATURE.

| Model | Supercapacitor | | Battery | |
|---|---|---|---|---|
| | MAE (%) | RMSE (%) | MAE (%) | RMSE (%) |
| UKF + PA [8] | 0.6300 | 0.7300 | 0.7900 | 0.9500 |
| ANN | 0.1201 | 0.5384 | 0.0838 | 0.6246 |
| NARXNN | **0.0146** | **0.0231** | **0.0239** | **0.0263** |

TABLE. 2. TIME CONSUMPTION COMPARISON FOR BATTERY SOC ESTIMATION.

| | UKF [11] | NARXNN | Percentage difference (%) |
|---|---|---|---|
| Time consumption (s) | 5.49 | 4.90 | **11.36** |

## III. RESULTS AND DISCUSSION

Three test experiments- constant current constant voltage (CC+CV) charge-discharge experiment under controlled and elevated temperatures and the urban dynamometer driving schedule (UDDS) condition experiment were performed to verify the accuracy and complexity of the proposed model in estimating SOC of battery and SC, and as well as validate its reliability, applicability, and robustness. The performance of the proposed model is evaluated based on the mean absolute error (MAE) and root mean square error (RMSE). Generally, the lower these values, the higher the estimation accuracy.

*A.  Test experiment 1*

Typical charging and discharging at temperature-controlled environment experiments were carried out on battery and SC and data were collected.

*1) SC SOC estimation*

The SC dataset used was obtained from Maxwell BCAP0025 T01 25 F supercapacitor cell with a rated voltage and capacitance of 2.7 V and 25 F respectively, cycled at constant charge and discharge current at ~25ºC with a sampling frequency of 1 Hz [15]. The SOC estimation of the SC is shown in Fig. 3 with NARXNN having MAE and RMSE of 0.0146% and 0.0231% respectively. This indicates a good agreement between the actual value and the estimated value.

*2) Battery SOC estimation*

The battery dataset used is the BEXEL datasets [16] extracted from lithium-ion batteries. This dataset consists of measured voltage, current, capacity, etc. Measurements were performed during charging under CC+CV profile and during constant current discharging. Cell#7 which has an initial capacity of 7.08 Ah and cut-off voltages of 2.5 V and 4.2 V was used in this experiment. Figure 4 shows the SOC estimation plot of the battery. The MAE and RMSE of the SOC estimation for the battery and SC are presented and compared with other state-of-the-art methods in Table 1. It can be seen that NARXNN performed better than the unscented kalman filter with parameter adaption (UKF + PA) and ANN models with MAE and RMSE of 0.0239% and 0.0263% respectively. A look at the time consumption of the NARXNN and the UKF model for battery SOC estimation in Table 2, shows that the proposed model had a time consumption of 4.9s which represents an 11.36% reduction in complexity. The results show that accurate, less complex, and robust SOC estimation can be obtained by NARXNN.

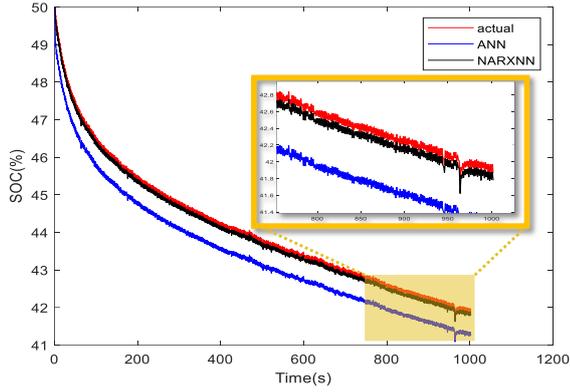

Fig. 5. SOC estimation results of ultracapacitor at different temperature of 45⁰C and an example interval (insert).

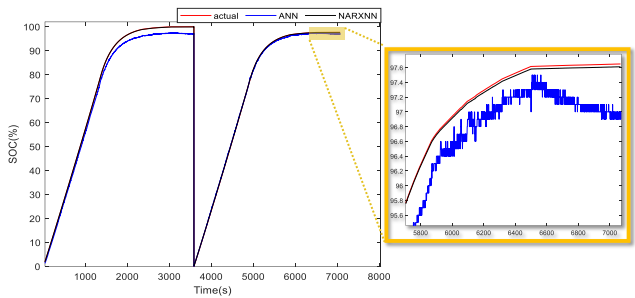

Fig. 6. SOC estimation results of battery at different temperature of 43⁰C and an example interval (insert).

### B. Test experiment 2

Another charging and discharging at elevated temperature experiments were carried out on battery and SC and data were collected.

#### 1) SC SOC estimation

The carbon electrodes-based commercial SCs (model HV, series 1 F 2.7 V from Eaton) were run through constant charge and discharge current of 20 mA at elevated ambient temperature - 45⁰C and data collected. The SOC estimation of the SC is shown in Fig. 5 with NARXNN having MAE and RMSE of 0.1526% and 0.3981% respectively. On the other hand, the ANN model had an MAE of 0.4041% and RMSE of 1.2920%. This further indicates that good SOC estimation result can be obtained with the proposed model for different temperature conditions.

#### 2) Battery SOC estimation

The NASA [17] battery dataset was used specifically cell#30 (1.66 Ah) with cut-off voltages of 2.5 V and 4.2 V. The cell's data were obtained from a charging and discharging experiment at elevated ambient temperature (43⁰C). The obtained SOC estimation results is displayed in Fig. 6. Based on the evaluation results of the ANN and NARXNN models shown in Table 3, it is observed that temperature has a significant impact on the SC and battery performance however, proposed NARXNN outperformed the ANN model with MAE value of 0.1781% against 0.9439% obtained by the ANN. Similar results was also achieved using the RMSE. The NARXNN improved the estimation results obtained by the ANN model by about 88.4%. The results further illustrate the accuracy and practicability of the proposed method.

TABLE. 3. PERFORMANCE OF SOC ESTIMATION UNDER CC+CV PROFILE AT DIFFERENT TEMPERATURE CONDITIONS.

| Model | Supercapacitor at 45⁰C | | Battery at 43⁰C | |
|---|---|---|---|---|
| | MAE (%) | RMSE (%) | MAE (%) | RMSE (%) |
| ANN | 0.4041 | 1.2920 | 0.9439 | 1.904 |
| NARXNN | **0.1526** | **0.3981** | **0.1781** | **0.2201** |

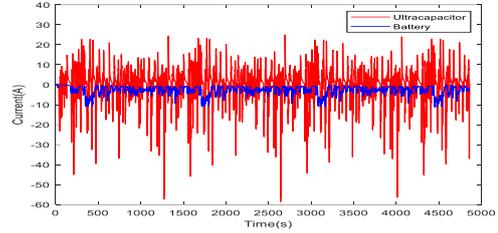

Fig. 7. Current profiles of battery and ultracapacitor under UDDS.

### C. Test experiment 3

The UDDS condition experiment was carried out to analyze the lithium-ion batteries (10 Ah) and ultracapacitors (3000 F) behavior at room temperature and data obtained [18]. The current profiles of the battery and ultracapacitor are shown in Fig. 7. This is a typical situation of complex and variable current charging and discharging operation as seen in today's HEVs. The SOC estimation of the ultracapacitor and battery is shown in Fig. 8 and 9 respectively. Similarly, Table 4 shows the MAE and RMSE of the proposed SOC method with other methods for batteries and ultracapacitor under UDDS profiles. The NARXNN had the best performance with MAE and RMSEof (0.0227%, 0.0465%) and (0.0016%, 0.0183%) for ultracapacitor and battery respectively. This indicates that NARXNN can overcome the nonlinear SOC behavior of battery and ultracapacitor with higher estimation accuracy. These results further illustrate the accuracy, practicability and robustness of the proposed method

## IV. CONCLUSION

This paper presented a novel deep learning method of SOC estimation for HESS using NARXNN. The method leverages the dynamic nonlinear property of the NARX model in overcoming the complex nonlinear nature of battery and SC SOC behavior. The proposed method was evaluated on actual datasets and results showed the effectiveness of the method in estimating the SOC of battery and SC for EV/HEV applications. As a part of future work, the proposed model would be a key part of an overall energy management system which operate in synchronization with other control algorithms of a HEV.

ACKNOWLEDGEMENT

The author wishes to appreciate Prof. M. Khanra of the Energy

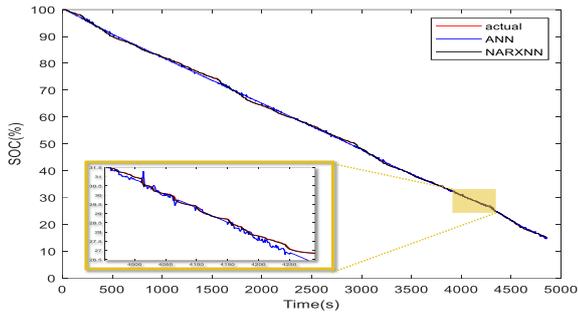

Fig. 8. SOC estimation results of ultracapacitor under UDDS and an example interval (insert).

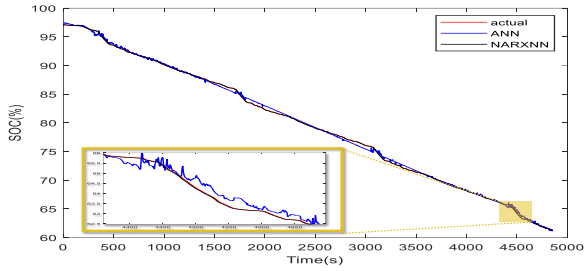

Fig. 9. SOC estimation results of battery under UDDS and an example interval (insert).

TABLE. 4. PERFORMANCE OF SOC ESTIMATION UNDER UDDS PROFILE.

| Model | Ultracapacitor | | Battery | |
|---|---|---|---|---|
| | MAE (%) | RMSE (%) | MAE (%) | RMSE (%) |
| UKF + PA [8] | 0.5000 | 0.6300 | 0.8300 | 2.0300 |
| ANN | 0.3337 | 0.5435 | 0.1278 | 0.7765 |
| NARXNN | **0.0227** | **0.0465** | **0.0016** | **0.0183** |